%% file: 0main.tex
\documentclass[conference]{IEEEtran}
\IEEEoverridecommandlockouts
\usepackage{subfigure}
\usepackage{amsmath,amssymb,amsfonts}
\usepackage{algorithmic}
\usepackage{graphicx}
\usepackage{textcomp}
\usepackage{xcolor}
\usepackage{xspace}
\usepackage{enumitem}
\usepackage{multirow}
\usepackage{multicol}
\usepackage{url}
\usepackage{cite}
\usepackage{marvosym} 
\usepackage{booktabs}

\newcommand{\n}[1]{{\color{black} #1}}

\newcommand{\name}{GPRec\xspace}

\def\BibTeX{{\rm B\kern-.05em{\sc i\kern-.025em b}\kern-.08em
    T\kern-.1667em\lower.7ex\hbox{E}\kern-.125emX}}

\makeatletter
\newcommand{\linebreakand}{%
  \end{@IEEEauthorhalign}
  \hfill\mbox{}\par
  \mbox{}\hfill\begin{@IEEEauthorhalign}
}
\makeatother
\begin{document}

\title{Bi-level User Modeling for Deep Recommenders
}

\author{
\IEEEauthorblockN{Yejing Wang}
\IEEEauthorblockA{
\textit{City University of Hong Kong} \\
Hong Kong, China \\
yejing.wang@my.cityu.edu.hk}
\and
\IEEEauthorblockN{Dong Xu}
\IEEEauthorblockA{
\textit{Xiaohongshu Inc}\\
Shanghai, China \\
xudong1@xiaohongshu.com}
\and
\IEEEauthorblockN{Xiangyu Zhao \Letter}
\IEEEauthorblockA{
\textit{City University of Hong Kong} \\
Hong Kong, China \\
xianzhao@cityu.edu.hk}
\linebreakand
\IEEEauthorblockN{Zhiren Mao}
\IEEEauthorblockA{
\textit{Xiaohongshu Inc}\\
Shanghai, China \\
maozhiren@xiaohongshu.com}
\and
\IEEEauthorblockN{Peng Xiang}
\IEEEauthorblockA{
\textit{Xiaohongshu Inc}\\
Shanghai, China \\
xiangpeng1@xiaohongshu.com}
\and
\IEEEauthorblockN{Ling Yan \Letter}
\IEEEauthorblockA{
\textit{Xiaohongshu Inc}\\
Shanghai, China \\
yanling@xiaohongshu.com}
\and
\IEEEauthorblockN{Yao Hu}
\IEEEauthorblockA{
\textit{Xiaohongshu Inc}\\
Shanghai, China \\
xiahou@xiaohongshu.com}
\linebreakand
\IEEEauthorblockN{Zijian Zhang}
\IEEEauthorblockA{
\textit{Jilin University}\\
Changchun, China \\
zhangzj2114@mails.jlu.edu.cn}
\and
\IEEEauthorblockN{Xuetao Wei}
\IEEEauthorblockA{
\textit{Southern University of Science and Technology}\\
Shenzhen, China \\
weixt@sustech.edu.cn}
\and
\IEEEauthorblockN{Qidong Liu}
\IEEEauthorblockA{\textit{Xi'an Jiaotong University} \\
\textit{City University of Hong Kong}\\
Xi'an, China \\
liuqidong@stu.xjtu.edu.cn
}
}

\maketitle

\begin{abstract}
Deep Recommender Systems (DRS) are essential for navigating the extensive data across various platforms in today's digital landscape. Current DRS models often treat all features equally and implement complex structures to enhance the capture of feature interactions. However, they may fail to recognize crucial user patterns due to not fully utilizing user-specific features for user modeling. Moreover, prevailing user modeling techniques concentrate exclusively on either the group or individual level, overlooking the potential insights from the unaddressed one. This oversight can miss shared group preferences or learn group patterns that conflict with individual preferences. To overcome these limitations, we introduce \textbf{\name}, a novel bi-level user modeling approach that substantially improves DRS. \name explicitly categorizes users into groups in a learnable manner and aligns them with corresponding group embeddings. We design the dual group embedding space to offer a diverse perspective on group preferences by contrasting positive and negative patterns. On the individual level, \name identifies personal preferences from ID-like features and refines the obtained individual representations to be independent of group ones, thereby providing a robust complement to the group-level modeling. We also present various strategies for the flexible integration of \name into various DRS models. Rigorous testing of \name on three public datasets has demonstrated significant improvements in recommendation quality. Additional experiments further explore crucial components of \name, its parameter sensitivity, and the group diversity. The implementation code is readily available online to facilitate future research and practical deployment: \url{https://github.com/Applied-Machine-Learning-Lab/GPRec}.
\end{abstract}

\input{1Intro}
\input{2Frame}

\input{3Exp}

\input{4Rel}
\input{5Con}

\bibliographystyle{IEEEtran}

\end{document}

%% file: 1Intro.tex
\section{Introduction}\label{sec:intro}
\n{With the advent of the information explosion, Deep Recommender Systems (DRS) have emerged as indispensable tools for sifting through data overload~\cite{liu2023multimodal,wang2023multi}. These systems are widely implemented across various online services~\cite{wang2023doctor,liu2023linrec,liu2023diffusion}, including shopping websites, content streaming platforms and social media networks, where they enhance the user experience by delivering personalized recommendations~\cite{liu2024large,li2022gromov}. 
Existing DRS models can be formally represented by $\mathcal{F}(\boldsymbol{x})$~\cite{yan2022apg}, where $\boldsymbol{x}$ encompasses features derived from users and items, i.e., $\boldsymbol{x} = [\boldsymbol{x}_u, \boldsymbol{x}_v]$, with $\mathcal{F}$ denoting the model architecture.

To improve the prediction capacity of DRS, considerable efforts have been devoted to designing advanced model architectures ($\mathcal{F}$) that better capture predictive feature interactions~\cite{lin2022adafs,lin2023autodenoise,li2023strec,jia2024erase}. Typical frameworks include linear structures \cite{zhang2016ffn,li2023automlp} 
, factorization machines \cite{rendle2010fm, blondel2016hofm, he2017nfm}, attention mechanism-based frameworks \cite{cheng2021SAM, xu2021destine, xiao2017afm}. 
Researchers have also developed advanced structures like cross networks for better feature crossing \cite{wang2017dcn, wang2023GDCN, qu2016pnn} and gating mechanisms \cite{jacobs1991moe} for better interaction fusion \cite{zhu2023finalnet, wang2023GDCN}. 
Nevertheless, these approaches often employ a uniform treatment across each feature field—whether user or item features—thus neglecting the crucial necessity of prioritizing user features. Such prioritization is vital for uncovering valuable user patterns that are specifically tailored to the personalization needs of both individuals and groups. This lack of emphasis on user features can result in suboptimal performance~\cite{yan2022apg,chang2023ppnet,ma2023JD}.

Consequently, integrating user modeling on user features ($\boldsymbol{x}_u$) into DRS—a relatively underexplored area—holds promise for improving prediction ability. Current user modeling methods are generally categorized into two levels: 
\begin{itemize}[leftmargin=*]
    \item \textbf{Individual Modeling:} These approaches usually 
    create specialized structures or parameters tailored for each user~\cite{yan2022apg,chang2023ppnet}, formally denoted as $\mathcal{F}_{u}(\boldsymbol{x})$. 
    \item \textbf{Group Modeling:} This level first divides users into groups according to user features, denoted as $g(\boldsymbol{x}_u)$ (ranging from $1$ to $G$ for $G$ groups). Then, different architectures and parameters are assigned to each group as $\mathcal{F}_{g(\boldsymbol{x}_u)}(\boldsymbol{x})$.  
    Compared with the individual level, group modeling generates coarse-grained structures, offering a more generalized but less personalized solution.
\end{itemize}
However, existing approaches tend to concentrate exclusively on one modeling level, thus neglecting the potential benefits of the complementary levels. 
On the one hand, only considering individual modeling~\cite{yan2022apg,chang2023ppnet} 
can obscure inter-user connections, preventing the system from fully leveraging collaborative knowledge and achieving suboptimal results. On the other hand, overemphasizing group patterns~\cite{ma2018mmoe,tang2020ple,wang2023single,sheng2021star,liu2023multi,ma2023JD} that may conflict with individual preferences can lead to discrepancies between personal interests and the characteristics of the assigned group, thereby potentially misguiding the DRS. For example, a user who prefers jazz music might be categorized into a broader music enthusiast group, resulting in inappropriate classical music recommendations. \textit{Therefore, there is a clear need for a novel user modeling method that effectively integrates both group and individual patterns to improve various DRS structures.}}
\begin{figure*}[t]
    \raggedleft
    \includegraphics[width=\linewidth]{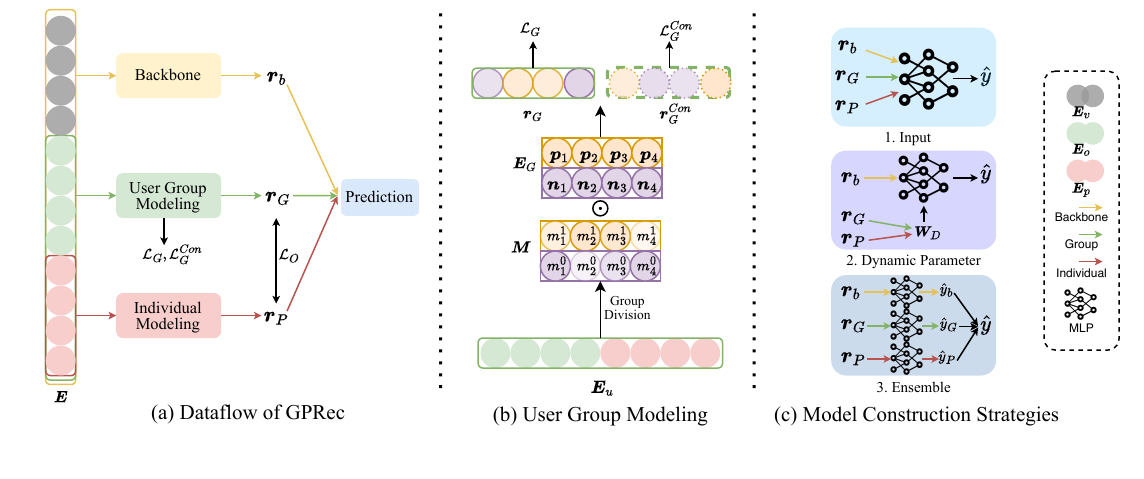}
    \vspace{-6mm}
    \caption{Framework overview.}
    \vspace{-1mm}
    \label{fig:Frame}
\end{figure*}

This paper introduces \textbf{\name}, a modular bi-level user modeling method, to improve \textbf{\underline{Rec}}ommendations with user \textbf{\underline{G}}roup modeling and enhanced \textbf{\underline{P}}ersonalization. \name effectively learns and utilizes both user group and individual representations, then adaptively integrates these two representations into DRS frameworks. Specifically, our approach first determines user groups according to user features and returns corresponding group representations. At the same time, individual representations are derived from user personal profiles. Then, we developed a series of model construction methods to integrate these representations into DRS, achieving the bi-level user modeling as $\mathcal{F}_{u,g(\boldsymbol{x}_u)}(\boldsymbol{x})$. 
A key challenge remains in refining the quality of group representations for more accurate recommendations. To address this limitation, we introduce dual embeddings for each group, coupled with the application of a contrastive loss. This design facilitates learning diverse group patterns, thereby significantly enhancing the quality of group representations. We summarize major contributions of this paprer as follows:
\begin{itemize}[leftmargin=*]
    \item To the best of our knowledge, we are the first to identify the importance of incorporating both group and individual preferences, and correspondingly propose the bi-level user modeling recommendation framework;
    \item We design dual group embedding space and contrastive loss for diverse group representations, along with the orthogonal loss to refine individual representations focusing on distinct personal preferences. 
    \item The proposed \name is a plug-in framework that can be flexibly applied to all DRS structures with various strategies; 
    \item We conduct extensive experiments on three public datasets to demonstrate the effectiveness of the proposed \name.
\end{itemize}

%% file: 2Frame.tex
\section{Notations}

As the understanding of DRS structures can vary significantly, this section elaborates on the specific DRS structure referenced in this paper, as well as commonly used notations and definitions. 

We structure the DRS into three modules: feature input, representation learning, and prediction. 
Specifically, an input sample $\boldsymbol{x}$ typically includes item features $\boldsymbol{x}_v$ and user features $\boldsymbol{x}_u$, formatted as one-hot vectors. We split user features into personal features ($\boldsymbol{x}_p$) and other features ($\boldsymbol{x}_o$) based on prior knowledge\footnote{In this paper, personal features are defined as those that potentially indicate user individual preferences, primarily encompassing attributes such as ‘UserID’. 
}. 
Thus, this paper represents $\boldsymbol{x}$ as $\boldsymbol{x} = [\boldsymbol{x}_p,\boldsymbol{x}_o,\boldsymbol{x}_v]$. The feature input module converts these one-hot vectors into low-dimensional dense vectors:
\begin{gather}
\boldsymbol{E}= [\boldsymbol{E}_p,\boldsymbol{E}_o,\boldsymbol{E}_v]= \boldsymbol{V}[\boldsymbol{x}_p,\boldsymbol{x}_o,\boldsymbol{x}_v]\label{eq:embed}
\end{gather}
Here, $\boldsymbol{V}$ represents the embedding table for all features. Equation~\eqref{eq:embed} maps the one-hot features $\boldsymbol{x}$ to their corresponding embeddings in $\boldsymbol{V}$, and returns the dense matrix $\boldsymbol{E}$.

Following the feature input is the representation learning, whose structure is determined by the chosen backbone model, including options like the cross network~\cite{wang2017dcn}, product network~\cite{qu2016pnn}, and other advanced architectures~\cite{wang2023GDCN,zhu2023finalnet,mao2023finalmlp}. This module transforms $\boldsymbol{E}$ into hidden states $\boldsymbol{r}_b$, which denote the representation learned from the backbone. Finally, $\boldsymbol{r}_b$ is input into the prediction module, typically a Multi-Layer Perceptron (MLP), to generate the output $\hat y_b$. 

\section{Framework}\label{sec:Frame}
This section provides an overview of \name, and elaborates on the bi-level user modeling, i.e., user group modeling and individual preference learning. Additionally, we discuss the model construction strategies to emphasize how \name can be conveniently combined with DRS models, and conclude with the optimization method.

\subsection{Overview}
Figure~\ref{fig:Frame} presents the \name framework with three components: Figure~\ref{fig:Frame} (a) depicts the overall data flow, Figure~\ref{fig:Frame} (b) focuses on user group modeling, and Figure~\ref{fig:Frame} (c) elaborates on the model construction strategies. 
In this figure, the gray balls represent the embeddings of item features ($\boldsymbol{E}_v$), the green balls represent the embeddings of other user features ($\boldsymbol{E}_o$), and the red balls denote the embeddings of user personal features ($\boldsymbol{E}_p$).
The data flow of the backbone model is indicated by yellow arrows. Additionally, green and red arrows highlight the \name's data flows associated with group modeling and individual preference learning.

The overall process of \name is visualized in Figure~\ref{fig:Frame} (a). We can see that the workflow of the backbone remains unaffected, indicating the modular design of \name. It processes all feature embeddings $\boldsymbol{E} = [\boldsymbol{E}_p,\boldsymbol{E}_o,\boldsymbol{E}_v]$ to produce the hidden state $\boldsymbol{r}_b$. \name's role involves user group modeling (Section~\ref{subsec:UGM}) and individual preference learning (Section \ref{subsec:Per}). The group modeling uses all user features $\boldsymbol{E}_u = [\boldsymbol{E}_p,\boldsymbol{E}_o]$ to generate the group representation $\boldsymbol{r}_G$, accompanied by two auxiliary losses $\mathcal{L}_G, \mathcal{L}^{Con}_{G}$. The individual modeling focuses on capturing the user's unique preferences $\boldsymbol{r}_P$ from personal embeddings $\boldsymbol{E}_p$ to enhance personalization, where an orthogonal loss is introduced, denoted as $\mathcal{L}_O$. Ultimately, the base representation $\boldsymbol{r}_b$, together with group and individual representations $\boldsymbol{r}_G,\boldsymbol{r}_P$, are integrated into the prediction module to generate the final prediction $\hat{y}$.

\subsection{User Group Modeling}\label{subsec:UGM}
This section details the user group modeling, including the group division method, the dual group embedding space, and the processes of obtaining user group representations $\boldsymbol{r}_G$ and two auxiliary losses.

The structure is visualized in Figure~\ref{fig:Frame} (b) for easier comprehension, which is exemplified by four user groups. \name first divides users into groups based on user embeddings ($\boldsymbol{E}_u$). The division is represented by binary masks ($\boldsymbol{M}$). The corresponding group representation ($\boldsymbol{r}_G$) is then obtained from the dual group embedding space ($\boldsymbol{E}_G$). Additionally, a contrasting group representation ($\boldsymbol{r}_G^{Con}$) is produced, utilizing the inverse of the group division result ($1-\boldsymbol{M}$). Two auxiliary losses ($\mathcal{L}_G$ and $\mathcal{L}^{Con}_{G}$) are computed during the group modeling. In the depicted example, the user is classified into Group\#2 and Group\#3 (denoted in orange) while not included in Group\#1 or Group\#4 (denoted in purple). Consequently, the user group representation $\boldsymbol{r}_G$ includes positive embeddings for Group\#2 and Group\#3 and negative embeddings for Group\#1 and Group\#4. And the contrasting representation $\boldsymbol{r}_G^{Con}$ is the reverse, i.e., including positive embeddings for Group\#1 and Group\#4 and negative embeddings for Group\#2 and Group\#3.

\textit{\textbf{Group Division. }}\n{
A straightforward method for group division is organizing groups based on explicit user attributes such as `Gender' and `Age' \cite{ma2018mmoe,tang2020ple,sheng2021star}. However, this attribute-centric approach has notable limitations due to its rigidity. It relies heavily on prior knowledge to select division criteria and binds users to fixed, unmodifiable groups post model construction. Another method used in existing work involves calculating cosine similarities between user embeddings and group embeddings. This method first computes the similarities between user embeddings and each group embedding. It then employs a Softmax operation over these results to determine users' connections to each group. Finally, it calculates a weighted sum of the group embeddings to represent the user group representations \cite{ma2023JD}. Nevertheless, this approach is limited to linear relationships and risks overfitting group embeddings to user embeddings, which can reduce diversity.

In response, we propose a learnable classifier, $g$, for group division, which can incorporate various deep learning structures. This classifier is designed to capture the non-linear dynamics within user embeddings $\boldsymbol{E}_u$, enabling a more nuanced and flexible division of user groups. Furthermore, $g$ helps to decouple the learning of group embeddings from user embeddings, thereby enhancing the diversity and effectiveness of groups.} This group division process is formulated as:
\begin{gather}
    \boldsymbol{S} = [s_{1}^1,\dots,s_{G}^1;s_{1}^0,\dots,s_{G}^0] = g(\boldsymbol{E}_u;\theta_g) \label{eq:gCls}
\end{gather}
where $\boldsymbol{S}\in \mathbb{R}^{2\times G}$ denotes the array of classification scores for $G$ groups. And $s_{i}^1$ reflects the classification score for assigning a user to group $i$, $s_{i}^0$ is the score for excluding a user from group $i$. $\theta_g$ denotes the learnable parameters of $g$.

\n{To achieve a more definitive group division for diverse group representation, we apply the Gumbel-Softmax technique \cite{maddison2017concrete,jang2016categorical}. Specifically, this technique converts the smooth scores in $\boldsymbol{S}$ into binary-like masks $\boldsymbol{M}$, approximating values close to $0$ or $1$~\cite{wang2022autofield}. This transformation is crucial because using smooth scores to fuse positive and negative group embeddings may lead to the blending of contradictory group patterns, potentially reducing their distinctiveness and impairing the efficiency.} $(m_{i}^1,m_{i}^0)$ in $\boldsymbol{M}$ are generated as:
\begin{gather}
m_{i}^1=\frac{\exp \left(\left(\log s_{i}^1+gn^{1}\right) / \tau\right)}{\exp \left(\left(\log s_{i}^1+gn^1\right) / \tau\right)+\exp \left(\left(\log s_{i}^0+gn^0\right) / \tau\right)} \label{eq:tau}\\
\text { where } gn^j = -\log \left(-\log \left(u^j\right)\right),\nonumber\\
u^j \sim \mathrm{Uniform}(0,1),\forall j \in [0,1]\nonumber\\
m_{i}^0=1-m_{i}^1
\end{gather}
In Equation~\eqref{eq:tau}, the temperature $\tau$ plays a crucial role in determining how closely the masks approximate 0 or 1, thereby potentially allowing us to adjust the level of differentiation between positive and negative group embeddings. Besides, ${\{gn^j\}}$ denotes a set of independent and identically distributed Gumbel noise, which boosts the training of $g$ and is not used during the inference. Note that users can be assigned to multiple positive groups, rather than only one of the group or the merge of all groups as existing works~\cite{ma2018mmoe,ma2023JD}.

 \textit{\textbf{Dual Group Embedding. }}\n{
 Existing group modeling methods typically assign a single embedding for each group, which poses challenges in efficient parameter utilization and capturing diverse group patterns. For example, with $G$ predefined group embeddings, the method can represent only $G$ distinct patterns. Furthermore, group embeddings are combined into a weighted sum for subsequent preference prediction, leading to a uniform learning direction for all group embeddings—the only variation being the weights. This can reduce the diversity of group patterns and diminish the effectiveness. In response, we design a dual group embedding space.}

Specifically, dual group embedding means assigning two distinct embeddings to each group: a positive group embedding and a negative one. The positive group embedding is designed to represent the preferences of users within the group, while the negative one is tailored to reflect a reverse inclination for users not belonging to the group. We anticipate a substantial divergence between the positive and negative embeddings of the same group, which might be crucial in enhancing the representation ability of dual group embeddings. For illustrative purposes, if a group's positive embedding identifies a preference for romance movies, we expect the negative embedding to depict an aversion to such movies or a preference for contrasting genres, such as thrillers, typically unpopular among users like romance movies. In the following text, the dual group embeddings are formulated as:
\begin{gather}
    \boldsymbol{E}_G = [\boldsymbol{p}_1,\dots,\boldsymbol{p}_G;\boldsymbol{n}_1,\dots,\boldsymbol{n}_G] \label{eq:Gemb}
\end{gather}
where $\boldsymbol{p}_i$ and $\boldsymbol{n}_i$ are positive and negative group embeddings for group $i$, respectively. And $G$ is the number of user groups. 

\n{Compared with the existing methods~\cite{ma2023JD}, the employment of dual group embedding exponentially enhances the representational scope of $G$ group embeddings from $G$ distinct patterns to $2^G$ by attributing either a positive or negative embedding of each group. 
In addition, the dual embedding space allows groups to learn user preferences from different directions. When users do not belong to specific groups, negative embeddings are assigned to learn their preferences. This arrangement enables positive embeddings to focus exclusively on the preferences of users who belong to the group.}

Finally, the group representation $\boldsymbol{r}_G$ is obtained by multiplying masks $\boldsymbol{M}$ and group embeddings $\boldsymbol{E}_G$:
\begin{gather}
    \boldsymbol{r}_G = \boldsymbol{M}\odot\boldsymbol{E}_G\overset{\underset{\mathrm{def}}{}}{=}[m_{1}^1\boldsymbol{p}_1+m_{1}^0\boldsymbol{n}_1,\dots,m_{G}^1\boldsymbol{p}_G+m_{G}^0\boldsymbol{n}_G] \label{eq:Ggen}
\end{gather}
\n{
In this equation, $G$ group patterns are concatenated, with each pattern represented as a weighted sum of positive and negative group embeddings. In practice, when $G$ is excessively large, summing up all group patterns serves as an alternative method to decrease the dimensionality of $\boldsymbol{r}_G$, thus enhancing computational efficiency.}

In order to effectively correlate group representation with user preferences, we integrate an auxiliary supervision task, predicting the preference for items (with embedding $\boldsymbol{E}_v$) based solely on $\boldsymbol{r}_G$:
\begin{gather}
    \hat y_1 = f_1(\boldsymbol{r}_G,\boldsymbol{E}_v;\theta_1)\label{eq:predG}\\
    \mathcal{L}_G = L(\hat y_1,y)\label{eq:LG_L}
\end{gather}
where $L$ is the loss function. $f_1$ is an MLP with parameter $\theta_1$, $\hat y_1$ is the prediction result, and $y$ is the ground truth. 

We expect positive and negative group embeddings to represent distinctly different user preferences. As a result, group embeddings should fail to accurately represent the user's preferences if the user is incorrectly classified. 
Consequently, we have integrated a contrastive loss to further refine the learning of group embeddings: 
\begin{gather}
      \boldsymbol{r}^{Con}_{G} = (1-\boldsymbol{M})\odot\boldsymbol{E}_G\label{eq:conGgen}\\
      \hat y_2 = f_1(\boldsymbol{r}^{Con}_{G},\boldsymbol{E}_v;\theta_1)\label{eq:predConG}\\
    \mathcal{L}^{Con}_{G} = L(\hat y_2,y)\label{eq:LConG_L}
\end{gather}
$f_1$ in Equation~\eqref{eq:predConG} is the same as in Equation~\eqref{eq:predG} but with $\theta_1$ fixed. This auxiliary task is specifically adopted to deepen the contrast between the positive and negative group embeddings. That is to say, we aim to maximize $\mathcal{L}^{Con}_{G}$ in Equation~\eqref{eq:LConG_L}.

\n{
\textit{\textbf{Discussion. }}Here, we summarize the differences between \name and the existing flexible user modeling method \cite{ma2023JD} to highlight our distinct contributions:
(\textbf{\textit{i}}) We employ a learnable deep classifier $g$, which is capable of capturing non-linear relationships between users and groups, as opposed to merely assigning users to groups based on linear similarities. And the result of the group division is extended to the shape of $\mathbb{R}^{2 \times G}$, which can representing a more nuanced group preference of users ($\mathbb{R}^{G}$ in \cite{ma2023JD} or $\mathbb{R}^{1}$ for multi-task/scenario methods~\cite{ma2018mmoe,sheng2021star}). 
(\textbf{\textit{ii}}) 
Along with the division result in $\mathbb{R}^{2 \times G}$, we design the dual embedding space for groups to denote contrasting preference. Compared with the single group embedding in \cite{ma2023JD}, our \name can learn more diverse group preference as validated in Section~\ref{sec:vis}. Besides, we introduce the Gumbel-Softmax technique and an additional contrastive loss $\mathcal{L}^{Con}_{G}$ to enhance the efficiency of dual embeddings.
(\textbf{\textit{iii}}) 
\name additionally considers individual preferences (Section~\ref{subsec:Per}) rather than merely group patterns. And we provide flexible model construction strategies (Section~\ref{sec:modelcons}), unlike focusing on dynamic weight applications in \cite{ma2023JD}.
}

\subsection{Individual Preference Learning}\label{subsec:Per}
This section is dedicated to illustrating the other level of user modeling, individual preference learning. 

Solely concentrating on user group patterns and neglecting individual preferences could be detrimental to users with heterogeneous tastes, thereby constraining the capacity of DRS to improve recommendation quality. To mitigate this limitation, we discern individual preferences.

Specifically, general user attributes like `Gender', `Age', and `Occupation' are widespread amongst many users, but features like `UserID' are unique. 
We hypothesize that these unique features harbor a deeper level of individual preference. This understanding guides us to prioritize id-like features as personal features, aiming to delve into the rich individual preference. We extract individual representations $\boldsymbol{r}_P$ from personal features $\boldsymbol{E}_p$ by an MLP $f_2$ with parameter $\theta_2$:
\begin{gather}
\boldsymbol{r}_P = f_2(\boldsymbol{E}_p;\theta_2)
\end{gather}

\n{However, there is a potential overlap between the group representations $\boldsymbol{r}_G$ and individual representations $\boldsymbol{r}_P$. For instance, a user identified as a sports fan within a group might exhibit similar traits in their personal features, leading to redundancy. This overlap could dilute the representation of other individual preferences. For example, although the user in the same example might also engage with music content, his individual preference could be overshadowed by his interest in sports. Given that sports preference is already captured in group representations, diluting this preference in individual preference could enhance the visibility of his interest in music, thereby improving the diversity of recommendations. To address this, we incorporate an orthogonal loss in \name, designed to minimize overlaps between group and individual representations, thus accentuating the unique aspects of individual preferences. The orthogonal loss can be expressed as:
}
\begin{gather}
    \mathcal{L}_O = \dfrac{\boldsymbol{r}_G\cdot \boldsymbol{r}_P}{||\boldsymbol{r}_G||_2||\boldsymbol{r}_P||_2}\label{eq:LO}
\end{gather}
where $||\cdot||_2$ indicates L2-norm. Equation~\eqref{eq:LO} computes the cosine similarity between $\boldsymbol{r}_G$ and $\boldsymbol{r}_P$, which is frequently utilized for decorrelation purposes~\cite{qi2022profairrec,chen2023biassurvey}.

\subsection{Model Construction}\label{sec:modelcons}
This section presents three model construction strategies combining our \name with DRS models. The strategies detailed in this section are illustrated in Figure~\ref{fig:Frame} (c).

\n{
Bi-level user modeling representations ($\boldsymbol{r}_G$ and $\boldsymbol{r}_P$) are readily combinable with existing DRS frameworks. 
Given the heterogeneous structures of DRS in real-world applications, a one-size-fits-all prediction model for GPRec is not universally effective.
Therefore, we demonstrate three model construction strategies as demonstrative examples: the input strategy, the dynamic parameter strategy, and the ensemble strategy. It is important to note that \name's design is inherently adaptable and can be integrated with any backbone, allowing additional strategies to be employed as needed.}


\textit{\textbf{Input Strategy.}} 
The input strategy is conceptually straightforward. This approach inputs the additional representations provided by \name to the prediction module ($f_y$) as:
\begin{gather}
    \hat y = f_y(\boldsymbol{r}_b,\boldsymbol{r}_G,\boldsymbol{r}_P;\theta_y) \label{eq:inputconstruct}
\end{gather}
where $\boldsymbol{r}_b$ signifies the base representation from the backbone. And $\theta_y$ denotes the trainable parameter of $f_y$.

\textit{\textbf{Dynamic Parameter (DP) Strategy.}}
Drawing inspiration from recent advancements in incorporating dynamic parameters into the prediction MLP of DRS models~\cite{yan2022apg,ma2023JD}, we explore the integration of \name using this technique:
\begin{gather}
    \boldsymbol{W}_{DP} = \mathrm{ReShape}(f_{DP}(\boldsymbol{r}_G,\boldsymbol{r}_P;\theta_{DP})) \label{eq:dpgenW}\\
    \hat y = f_y(\boldsymbol{r}_b;\theta_y+\boldsymbol{W}_{DP}) \label{eq:dpconstruct}
\end{gather}
Equation~\eqref{eq:dpgenW} dynamically generates parameters $\boldsymbol{W}_{DP}$ for the prediction MLP $f_y$ based on $\boldsymbol{r}_G$ and $\boldsymbol{r}_P$ through an additional MLP ($f_{DP}$). Then, Equation~\eqref{eq:dpconstruct} makes predictions based on $\boldsymbol{r}_b$ and prediction MLP with the parameter $\theta_y+\boldsymbol{W}_{DP}$. 

\textit{\textbf{Ensemble Strategy.}} 
Motivated by the effectiveness of ensemble structures in DeepFM~\cite{guo2017deepfm} and WD~\cite{cheng2016wd}, we propose the ensemble strategy for combining \name with DRS models. This approach separately generates three predictions based on $\boldsymbol{r}_b$, $\boldsymbol{r}_G$, and $\boldsymbol{r}_P$, respectively, and then aggregates them:
\begin{gather}
\hat y_b = f_b(\boldsymbol{r}_b;\theta_b),
\hat y_G = f_G(\boldsymbol{r}_G;\theta_G),
\hat y_P = f_P(\boldsymbol{r}_P;\theta_P)\label{eq:ensP}\\
\hat y = (\hat y_b+\hat y_G+\hat y_P)/3\label{eq:ensconstruct}
\end{gather}
In this strategy, the final prediction $\hat y$ is obtained by averaging the ensemble of the three predictors. Other ensemble methods, such as voting and weighted averaging, are also applicable. 

\begin{table}[t]
\centering
\caption{Dataset statistics.}
 \vspace{-2mm}
\label{tab:dataset}
\resizebox{\columnwidth}{!}{%
\begin{tabular}{@{}ccccccc@{}}
\toprule
       & \# User & \# Item   & \# Interaction & \# User  & \# Total \\ \midrule
ML1M   & 6,040   & 3,706     & 1,000,209      & 5              & 8        \\
TenRec & 999,447 & 2,310,087 & 120,342,306    & 3             & 6        \\
KuaiRand & 27,077 & 7,551 & 1,436,609    & 30               & 89        \\
\bottomrule
\end{tabular}%
}
\vspace{-1mm}
\end{table}

\subsection{Optimization}
By integrating \name with backbone DRS, we can determine the final prediction $\hat y$ as Equation~\eqref{eq:inputconstruct}, Equation~\eqref{eq:dpconstruct}, or Equation~\eqref{eq:ensconstruct}. Then, the primary optimization objective is:
\begin{gather}
    \mathcal{L}_{major} = L(\hat y, y) \label{eq:LMajor}
\end{gather}
The loss function $L$ utilized in Equation~\eqref{eq:LConG_L}, Equation~\eqref{eq:LG_L}, and Equation~\eqref{eq:LMajor} is consistent with the specific recommendation task. For instance, for click-through rate prediction tasks, $L$ is defined as the binary cross-entropy loss (BCE), mathematically expressed as $L(\hat y,y) = y \cdot \log \hat y+\left(1-y\right) \cdot \log \left(1-\hat y\right)$. 

The overall optimization objective is formulated as:
\begin{gather}
    \mathcal{L} = \mathcal{L}_{major} + \lambda_1\mathcal{L}_G - \lambda_2\mathcal{L}^{Con}_{G} + \lambda_3\mathcal{L}_O \label{eq:ovopt}
\end{gather}
$\mathcal{L}_{major}$ is the main loss from Equation~\eqref{eq:LMajor}. $\mathcal{L}_G,\mathcal{L}^{Con}_{G}$, and $\mathcal{L}_O$ are auxiliary losses from previous sections. 
And $\lambda_1,\lambda_2,\lambda_3$ are hyper-parameters set to balance the training objectives. Finally, \name is an end-to-end system that enables the direct application of the gradient descent strategy.

%% file: 3Exp.tex
\begin{table*}[ht]
\centering
\caption{Performance comparison with baselines.}
\label{tab:OV}
\begin{tabular}{@{}cc|ccccccc@{}}
\toprule
\multicolumn{2}{c|}{Feature Interaction}            & MLP  & DeepFM & DCN & GDCN & FinalMLP    & DESTINE &GPRec(GDCN) \\\midrule
\multicolumn{1}{c|}{\multirow{2}{*}{ML1M}}     & AUC$\uparrow$     & 0.8081 & 0.8044& 0.8087 & \underline{0.8149} & 0.8101  & 0.8082 & \textbf{0.8161*} \\
\multicolumn{1}{c|}{}                            & LogLoss$\downarrow$ & 0.5407 & 0.5339 & 0.5326 & \underline{0.5196} & 0.5217 & 0.5314 & \textbf{0.5167*} \\\midrule
\multicolumn{1}{c|}{\multirow{2}{*}{TenRec}}   & AUC$\uparrow$     & 0.9126& 0.9134 & 0.9139 & 0.9140 & 0.9142  & \underline{0.9152} & \textbf{0.9180*}\\
\multicolumn{1}{c|}{}                            & LogLoss$\downarrow$ & 0.3479& 0.3281 & 0.3265 & 0.3324 & 0.3284  & \textbf{0.3222*} & \underline{0.3231} \\\midrule
\multicolumn{1}{c|}{\multirow{2}{*}{KuaiRand}} & AUC$\uparrow$     & 0.7530& 0.7403  & 0.7533 & \underline{0.7564} & 0.7553 & 0.7544 & \textbf{0.7573*} \\
\multicolumn{1}{c|}{}                            & LogLoss$\downarrow$ & 0.5857& 0.6120 & 0.5987 & \textbf{0.5832} & 0.5837  & 0.5854 & \underline{0.5835} \\
                          \toprule
\multicolumn{2}{c|}{User Modeling} & MMoE & PLE & STAR & APG      & PEPNet & DGPM    & GPRec(MLP)  \\\midrule
\multicolumn{1}{c|}{\multirow{2}{*}{ML1M}}     & AUC$\uparrow$     & 0.8081 & 0.8088 & \underline{0.8095} & 0.8090 & 0.8082 & 0.8086 & \textbf{0.8141*} \\
\multicolumn{1}{c|}{}                            & LogLoss$\downarrow$ & 0.5234 & 0.5226 & 0.5225 & \underline{0.5214} & 0.5280 & 0.5350 & \textbf{0.5181*} \\\midrule
\multicolumn{1}{c|}{\multirow{2}{*}{TenRec}}   & AUC$\uparrow$     & 0.9134 & 0.9150 & 0.9161 & 0.9136 & \textbf{0.9165} & 0.9146 & \underline{0.9163} \\
\multicolumn{1}{c|}{}                            & LogLoss$\downarrow$ & 0.3129 & \textbf{0.3090*} & \underline{0.3105} & 0.3320 & 0.3369 & 0.3235 & 0.3193 \\\midrule
\multicolumn{1}{c|}{\multirow{2}{*}{KuaiRand}} & AUC$\uparrow$     & 0.7523 & 0.7521 & 0.7517 & 0.7560 & \underline{0.7563} & 0.7540 & \textbf{0.7565} \\
\multicolumn{1}{c|}{}                            & LogLoss$\downarrow$ & 0.5877 & 0.5881 & 0.5876 & 0.5832 & \textbf{0.5827} & 0.5853 & \underline{0.5830}
                        \\ \bottomrule
\end{tabular}
\\``\textbf{{\Large *}}'' indicates the statistically significant improvements. (i.e., one-sided t-test with $p<0.05$) over the runner-up method.
\end{table*}

\section{Experiment}\label{sec:exp}
In this section, we provide detailed descriptions of our experimental results to address the following research questions:

\begin{itemize}[leftmargin=*]
    \item \textbf{RQ1: }How does \name perform compared with baselines?
    \item \n{\textbf{RQ2: } Is \name compatible with various DRS backbones?}
    \item \textbf{RQ3: }What is the impact of the core components?
    \item \textbf{RQ4: }How do hyper-parameters affect the effectiveness of 
 the proposed \name?
    \item \textbf{RQ5: }Can \name successfully learn contrasting and diverse group patterns as expected?
\end{itemize}

\subsection{Experiment Settings}
This section provides details of our experimental setup, including datasets, evaluation metrics, baselines, and implementation details.

\subsubsection{Dataset} We validate the effectiveness of \name on three public datasets:

\begin{itemize}[leftmargin=*]
\item \textbf{ML1M\footnote{https://grouplens.org/datasets/movielens/1m/}}, short for MovieLens 1M, a benchmark dataset commonly used in DRS research. It comprises 5 user features, with `userid' serving as the only user personal feature. 
\item \textbf{TenRec\footnote{https://github.com/yuangh-x/2022-NIPS-Tenrec}}, a large-scale dataset recently released by Tencent~\cite{yuan2022tenrec} collected from their browsers. We utilize the sampled version designed for the CTR prediction task as described in the original study. This dataset contains 3 user features, with `userid' as the unique personal feature.
\item \textbf{KuaiRand\footnote{https://kuairand.com/}}, collected from KuaiShou. This dataset contains 89 features, where 30 are user features. We adopt the pure version as advised by the literature~\cite{gao2022kuairand}.
\end{itemize} 
The detailed statistics are presented in Table~\ref{tab:dataset}, which includes the number of users, items, interactions, user features, and the total number of features. 
For all datasets, we randomly allocate $80\%$ for training, $10\%$ for validation, and the remaining $10\%$ for testing. 
\subsubsection{Evaluation Metric} We conduct the Click-Through Rate (CTR) prediction task on all datasets and assess the overall performance using the AUC score (Area Under the ROC Curve) and LogLoss (logarithmic loss). A larger AUC score or a lower LogLoss value at the 0.001 level can indicate a significant improvement \cite{guo2017deepfm}. Additionally, we employ the single-sided t-test~\cite{pfanzagl1996studies}  to validate the superiority of the best-performing methods over the second-best.
\subsubsection{Baselines} We compare \name with three sorts of methods:
\n{ (1) \textbf{Feature interaction modeling}: MLP~\cite{zhang2016ffn}, DeepFM~\cite{guo2017deepfm}, DCN~\cite{wang2017dcn}, GDCN~\cite{wang2023GDCN}, FinalMLP~\cite{mao2023finalmlp}, and DESTINE~\cite{xu2021destine}; (2) \textbf{Group modeling}: MMoE~\cite{ma2018mmoe}, PLE~\cite{tang2020ple}, STAR~\cite{sheng2021star}, and DGPM~\cite{ma2023JD}; (3) \textbf{Individual modeling}: APG~\cite{yan2022apg} and PEPNet~\cite{chang2023ppnet}. 
}

\subsubsection{Implementation Details}
For the baseline implementation, we utilized three public GitHub libraries\footnote{\url{https://github.com/rixwew/pytorch-fm}}\footnote{ \url{https://github.com/reczoo/FuxiCTR}}\footnote{\url{https://github.com/easezyc/Multitask-Recommendation-Library}}. The embedding size, both for DRS in Equation~\eqref{eq:embed} and for \name's group embeddings in Equation~\eqref{eq:Gemb}, is set to 16. The dimensions of the MLPs used for prediction purposes, including $f_1$ for group embedding learning (Section~\ref{subsec:UGM}), as well as $f_y,f_b,f_G,f_P$ for model construction (Section~\ref{sec:modelcons}), are configured as $[64,32,16,1]$. The MLP $F_G$, serving as the group classifier in Equation~\eqref{eq:gCls}, has dimensions of $[64,32,2]$ for each group, with the total number of groups (G) set at 60. For Equation~\eqref{eq:ovopt}, the parameters are $\lambda_1=1.0,\lambda_2=0.001,\lambda_3=0.0001$, and $\tau = 0.5$ as per Equation~\eqref{eq:tau}. 

\begin{table*}[]
\centering
\caption{Compatibility test of \name.}
\label{tab:my}
\begin{tabular}{@{}c|c|ccc|ccc|ccc|ccc@{}}\toprule
\multirow{2}{*}{Dataset} & \multirow{2}{*}{Metric} & \multicolumn{3}{c|}{DCN} & \multicolumn{3}{c|}{GDCN} & \multicolumn{3}{c|}{DESTINE} & \multicolumn{3}{c}{FinalMLP} \\\cmidrule(l){3-14}
         &         & Base   & DGPM   & GPRec  & Base   & DGPM   & GPRec  & Base   & DGPM   & GPRec   & Base   & DGPM & GPRec   \\\midrule
\multirow{2}{*}{ML1M}    & AUC$\uparrow$     & 0.8087 & 0.8095 & \textbf{0.8135*} & 0.8149 & 0.8150 & \textbf{0.8161*} & 0.8082 & 0.8095 & \textbf{0.8126*} & 0.8101 & -    & \textbf{0.8136*} \\
         & LogLoss$\downarrow$ & 0.5326 & 0.5352 & \textbf{0.5220*} & 0.5196 & 0.5181 & \textbf{0.5167*} & 0.5314 & 0.5282 & \textbf{0.5207*}  & 0.5217 & -    & \textbf{0.5195*}  \\\midrule
\multirow{2}{*}{TenRec}  & AUC$\uparrow$     & 0.9139 & 0.9146 & \textbf{0.9161*} & 0.9140 & 0.9145 & \textbf{0.9180*} & 0.9152 & 0.9152 & \textbf{0.9179*}  & 0.9142 & -    & \textbf{0.9173*}  \\
         & LogLoss$\downarrow$ & 0.3265 & 0.3244 & \textbf{0.3149*} & 0.3324 & 0.3301 & \textbf{0.3231*} & \textbf{0.3222} & 0.3227 & 0.3269  & \textbf{0.3284} & -    & 0.3292  \\\midrule
\multirow{2}{*}{KuaiRand} & AUC$\uparrow$     & 0.7533 & 0.7529 & \textbf{0.7553*} & 0.7564 & 0.7566 & \textbf{0.7573*} & 0.7544 & 0.7533 & \textbf{0.7552*} & 0.7553 & -    & \textbf{0.7564*} \\
         & LogLoss$\downarrow$ & 0.5987 & 0.5861 & \textbf{0.5837*} & 0.5832 & \textbf{0.5829} & 0.5835 & 0.5854 & 0.5860 & \textbf{0.5836} & 0.5837 & -    & \textbf{0.5832} \\ \bottomrule
\end{tabular}
\\``\textbf{{\Large *}}'' indicates the statistically significant improvements. (i.e., one-sided t-test with $p<0.05$) over the runner-up method.
\end{table*}

\subsection{Overall Performance (RQ1\&RQ2)}\label{sec:ov}

\n{
In this section, we conduct a comprehensive evaluation of \name on three public datasets, comparing its performance against baseline models and examining its compatibility with various DRS backbones. To ensure the reliability of our findings, each value is the average of 5 replicated tests with different seeds. 
For our \name, we present only the results of the best-performing model construction.

We display the \name with GDCN and MLP backbones in Table~\ref{tab:OV}, demonstrating the optimal performance and enabling a fair comparison with MLP-based baselines, respectively. To test the compatibility of \name, we also provide results with other backbones and compare these against the most similar baseline DGPM in Table~\ref{tab:my}. The results table marks the highest-performing method in bold, and the runner-up method is underlined. We conclude that:

\begin{itemize}[leftmargin=*]
\item 
From the upper half of Table~\ref{tab:OV}, which presents the results for `Feature Interaction' methods, we observe that advanced architectures (GDCN, FinalMLP, and DESTINE) significantly outperform classical models (MLP, DeepFM, DCN). This underscores the effectiveness of developing complex architectures to enhance recommendation quality. When integrated with the GDCN backbone, \name outperforms all baselines. Meanwhile, \name with the MLP backbone also demonstrates competitive results, as depicted in the last column of the lower half of Table~\ref{tab:OV}, surpassing most baselines except for GDCN. These findings highlight the significant improvements achieved by \name through integrating bi-level user modeling.




\item 
The lower half of Table~\ref{tab:OV} includes the results of `User Modeling' methods, all of which primarily comprise MLP components. It is observed that all baselines generally outperform the MLP (the first column in the upper half), validating the effectiveness of considering either user group patterns (MMoE, PLE, STAR, DGPM) or individual patterns (APG, PEPNet). However, methods that require predefined criteria for user division (MMoE, PLE, STAR) fail to enhance performance on KuaiRand. This underperformance can be attributed to inappropriate division criteria. Due to the lack of the `Gender' attribute (used for user division in ML1M and TenRec), we use `User Active Degree' as an alternative in KuaiRand, which may not be as effective. In contrast, the group modeling method with flexible group division (DGPM) demonstrates stability across datasets. This underscores the advantages of implementing flexible group division over rigid predefined groups. Additionally, user individual modeling methods prove to be very effective, highlighting the importance of recognizing individual patterns. With the integration of both group and individual patterns, our \name\ (MLP) achieves optimal performance among all MLP-based methods.

\item 
In Table~\ref{tab:my}, where `Base' indicates the results for backbone models without any additional structures, we observe that \name consistently enhances the backbone model and outperforms DGPM across all datasets and backbones. 
Notably, DGPM fails to integrate with FinalMLP, indicated with `-' in the table, whose prediction layer is not suitable for dynamic parameter integration—the sole construction strategy offered by DGPM. These results underscore the modular and adaptive design of \name, which allows it to seamlessly integrate with a variety of backbone architectures, thereby consistently enhancing their performance.
\end{itemize}
To sum up, the bi-level user modeling of \name contributes significantly to enhancing recommendation quality across various datasets and backbones. The plug-in design and flexible construction strategies enable it to seamlessly integrate with any backbones. Consequently, the proposed \name can serve as a convenient and effective solution for most DRS to improve their performance.

}


\begin{figure}[t]
    \centering
    \includegraphics[width=\linewidth]{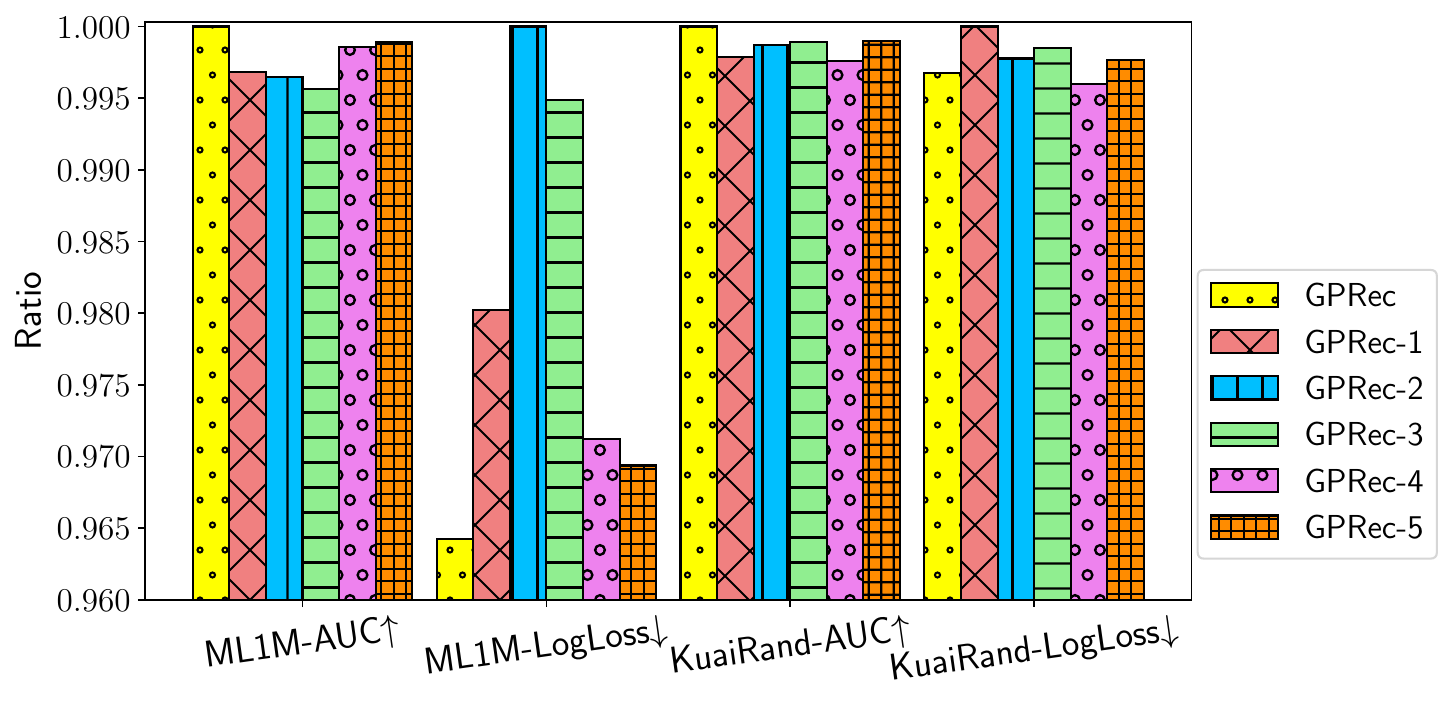}
    \vspace{-6mm}
    \caption{Ablation study for \name on ML1M and KuaiRand with MLP.}
    \label{fig:abl}
    \vspace{-3mm}
\end{figure}
 
\subsection{Ablation Study (RQ3)}\label{sec:abl}
In this section, we execute experiments to ascertain the significance of key components within \name. We introduce five variants to discern the contributions of dual group embedding ($\boldsymbol{E}_G$), individual preference ($\boldsymbol{r}_P$), and three auxiliary losses in Equation~\eqref{eq:ovopt} ($\mathcal{L}_G$,$\mathcal{L}_{ConG}$,$\mathcal{L}_O$):

\n{\begin{itemize}[leftmargin=*]
\item \textbf{\name -1:} This variant replaces dual embeddings with a single embedding for each group, correspondingly altering the group division method (Equation~\eqref{eq:gCls}-Equation\eqref{eq:Ggen}). Specifically, the dual embedding for each group is replaced by a single embedding, and Softmax is employed on $\boldsymbol{S}$ to classify users, similar to the approach used in the baseline DGPM~\cite{ma2023JD}. This modification is intended to evaluate the effectiveness of the dual group embedding ($\boldsymbol{E}_G$), alongside the group division method proposed.
\item \textbf{\name -2:} To assess the impact of individual representations $\boldsymbol{r}_P$ in Section~\ref{subsec:Per}, this version omits $\boldsymbol{r}_P$ entirely, including its involvement in $\mathcal{L}_O$ from Equation~\eqref{eq:ovopt} and its use in predicting $\hat{y}$ in Section~\ref{sec:modelcons} (e.g., Equation~\eqref{eq:dpconstruct}).
\item \textbf{\name -3, \name -4, and \name -5:} These three variants respectively remove $\mathcal{L}_G$, $\mathcal{L}_{ConG}$, $\mathcal{L}_O$ in Equation~\eqref{eq:ovopt} to validate the effectiveness of each auxiliary loss.
Each removal is designed to validate the effectiveness of each auxiliary loss in enhancing the overall model performance, enabling a focused analysis of the contribution of each component to \name.
\end{itemize}}
The performance of these variants was tested on ML1M and KuaiRand with MLP backbone. The results are illustrated in Figure~\ref{fig:abl}, which displays the ratio of performance (AUC, LogLoss) between the variants and the original \name. We can observe that:
\begin{itemize} [leftmargin=*]
    \item \name-1's underperformance highlights the critical role of the dual group embedding and corresponding group division method in learning effective group representations.
    \item The variant \name-2, which omits $\boldsymbol{r}_P$, records the high LogLoss on both datasets, indicating poor performance. This outcome strongly supports our assertion regarding the crucial role of individual preferences in enhancing DRS.
    \item \n{The underperformance of \name-3, \name-4, and \name-5 compared to the original \name demonstrates that the omission of any auxiliary loss can significantly degrade model performance. Notably, \name-3 and \name-4 exhibit the lowest performance on ML1M and KuaiRand, respectively. This underscores the importance of $\mathcal{L}_G$ and $\mathcal{L}_{ConG}$ in learning representative group embeddings, as expounded in Section~\ref{subsec:UGM}. Furthermore, as anticipated in Section~\ref{subsec:Per}, incorporating an orthogonal loss slightly improves individual modeling and benefits recommendations.}
\end{itemize}
In summary, this experiment supports the rationale and the effectiveness of essential components in \name, including dual group embeddings, individual preference modeling, and three auxiliary losses.

\subsection{Hyper-Parameter Study (RQ4)}
\begin{figure}[t]
 \centering
 {\subfigure{\includegraphics[width=0.49\linewidth]{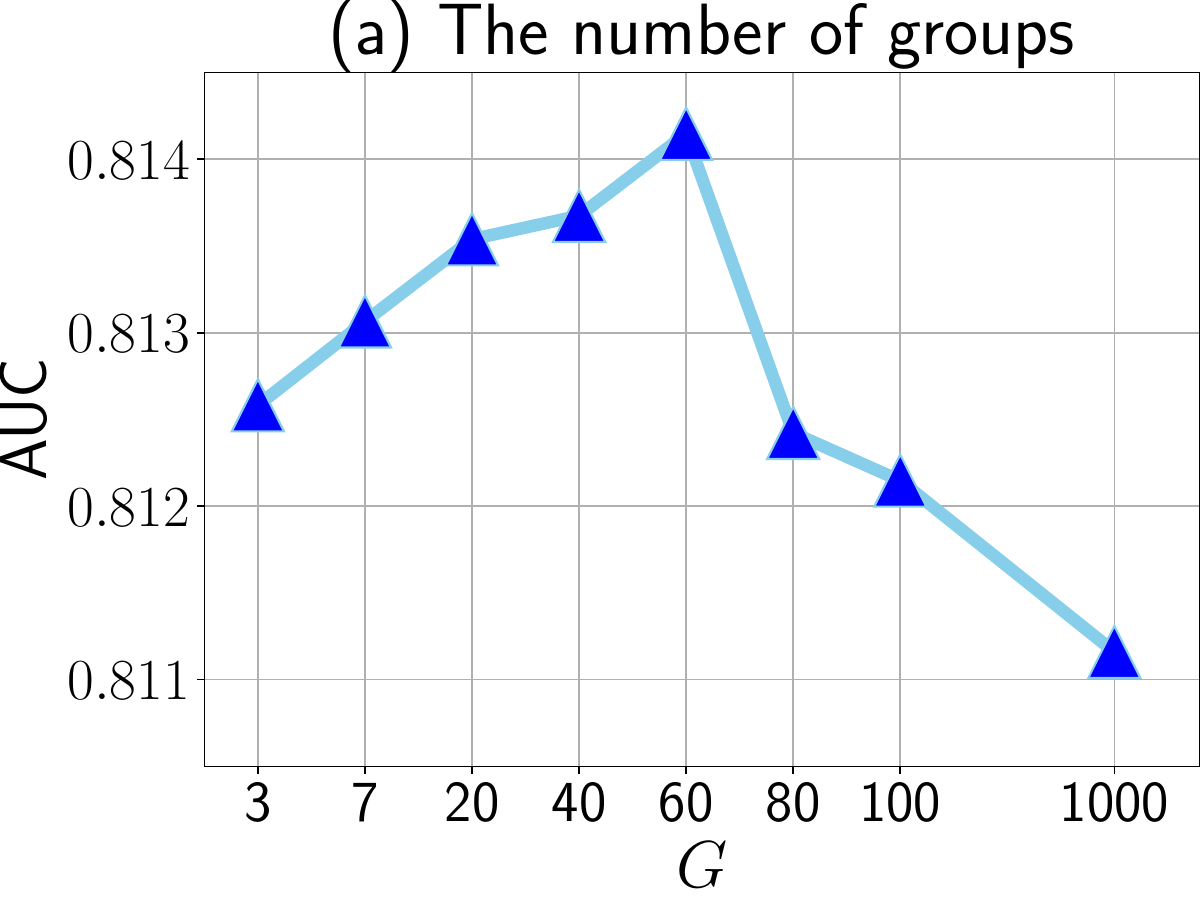}}}
 {\subfigure{\includegraphics[width=0.49\linewidth]{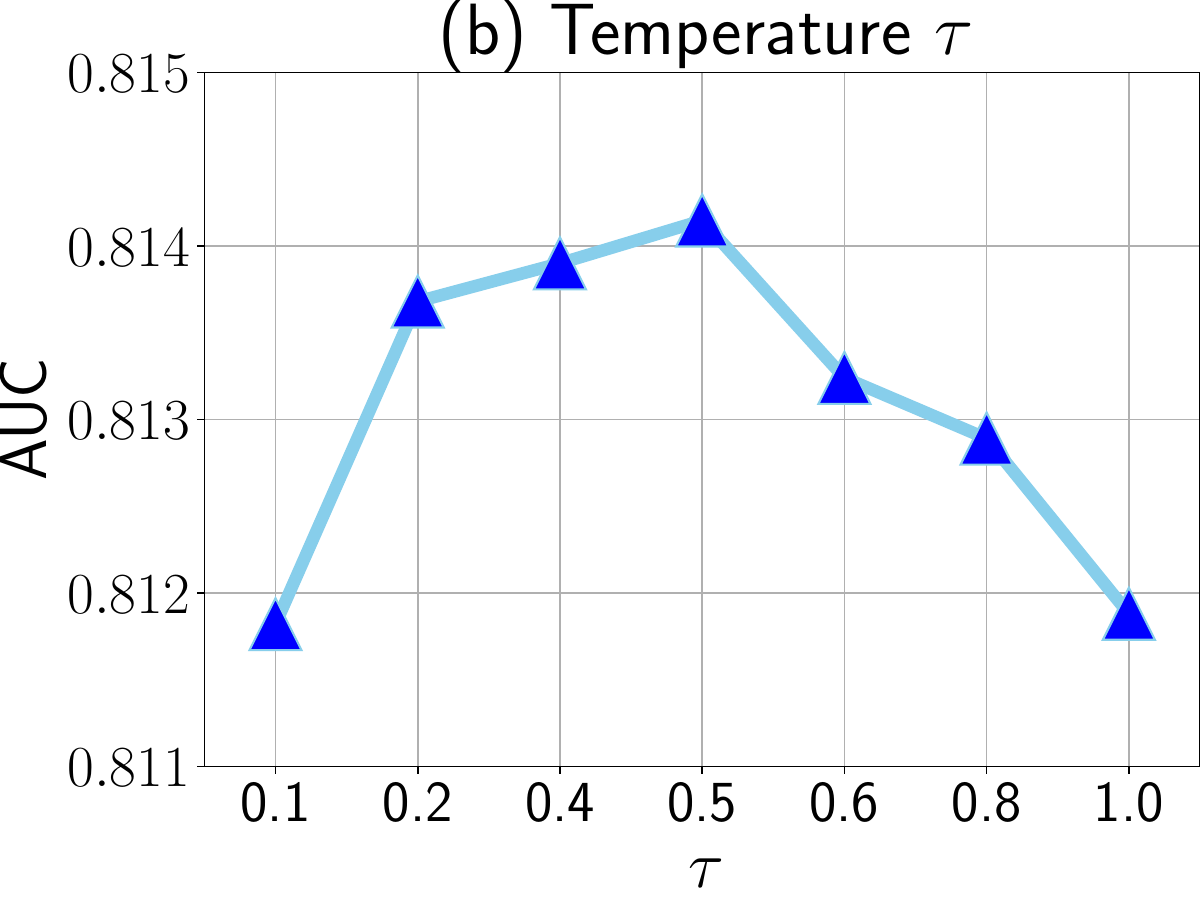}}}
 \vspace{-7mm}
 \caption{Parameter analysis on ML1M with MLP.}
 \label{fig:param}
\vspace{-3mm}
\end{figure}

In this section, we explore the influence of crucial parameters on \name. Our focus is primarily on the number of groups $G$ for $\boldsymbol{E}_G$ and the temperature parameter $\tau$ from Equation~\eqref{eq:tau}. We test the AUC score on the ML1M dataset with the MLP backbone. Parameters are altered one at a time, with $G$ varying among [3, 7, 20, 40, 60, 80, 100, 1000] and $\tau$ tested at values [0.1, 0.2, 0.4, 0.5, 0.6, 0.8, 1]. The results are presented in Figure~\ref{fig:param}. Notably, the AUC for the base MLP model, which stands at 0.8081, is omitted in Figure~\ref{fig:param} due to its comparatively low value. Our observations include:
\begin{itemize}[leftmargin=*]
    \item From Figure~\ref{fig:param} (a), the AUC score initially improves with increasing $G$. However, beyond a certain threshold, further increments in $G$ cease to enhance AUC. The reason is that \name initially benefits from more fine-grained group division, and an excessive number of groups can lead to undertraining of group embeddings.
    \item \n{Figure~\ref{fig:param} (b) demonstrates a similar trend with $\tau$. A highly definitive group division (small $\tau$) can confine users to static choices during the training, making the optimization stuck at the suboptimal state. While overly flexible division (large $\tau$) blurs positive and negative group patterns, impairing the learning of contrastive and predictive group embeddings.
}
    \item The optimal configuration is $G=60,\tau=0.5$. This setting was selected for overall performance evaluation in Section~\ref{sec:ov}. Remarkably, \name consistently surpasses the base model in AUC across all parameter settings, demonstrating its stability.
\end{itemize}

\subsection{Model Visualization (RQ5)}\label{sec:vis}
In this section, we examine group embedding similarities of \name to evaluate whether our expectations are fulfilled. 

Figure~\ref{fig:sim} displays the group embedding similarities obtained by \name on ML1M with MLP backbone ($G=60$). For clarity, we visually represent the cosine similarities between five randomly selected groups. The results indicate that \name successfully learns contrasting group embeddings, evidenced by the notably low similarities between negative and positive embedding pairs (the diagonal of the top right and bottom left quadrants, e.g., 0.03 for $\boldsymbol{p}_2$ and $\boldsymbol{n}_2$). Besides, the heatmap predominantly shows light green, with a maximum similarity of 0.6 between $\boldsymbol{n}_3$ and $\boldsymbol{n}_4$ (two negative group embeddings), suggesting a robust representation capability of group embeddings in \name. In comparison, group embeddings in DGPM~\cite{ma2023JD} exhibit an average similarity of 0.9. This finding aligns with our expectations in Section~\ref{subsec:UGM}, where we apply Gumbel-Softmax and design dual embedding space to obtain diverse representations.

%% file: 4Rel.tex
\section{Related Work}\label{sec:rel}
The proposed framework aims to improve DRS through bi-level user modeling. Consequently, this section discusses the related work from two perspectives: DRS architectures and user modeling methods. 

\textbf{DRS Architectures}
The architecture of DRS significantly impacts the performance, making it a popular area of focus in both industry and academia. Basic DRS architectures include linear structures, feature factorization machines (FM) \cite{rendle2010fm}, and attention mechanisms \cite{vaswani2017attention}. Other structures found effective for feature crossing or fusing interactions include cross networks \cite{wang2017dcn}, product networks \cite{qu2016pnn}, and gating mechanisms \cite{jacobs1991moe}. Combining these components has led to the development of numerous popular architectures, such as MLP \cite{zhang2016ffn}, DeepFM \cite{guo2017deepfm}, and Wide \& Deep (WD) \cite{cheng2016wd}. More advanced DRS architectures have been developed in recent years based on these foundational components \cite{xu2021destine, cheng2021SAM, mao2023finalmlp, zhu2023finalnet, wang2023GDCN}. For example, GDCN~\cite{wang2023GDCN} designs the gating mechanism for crossed features based on DCN~\cite{wang2017dcn}, significantly improving the recommendation quality and interpretability. FinalMLP~\cite{mao2023finalmlp} invises a dual stream network with bi-linear fusion as the prediction layer for explicit feature interaction modeling. While DESTINE~\cite{xu2021destine} proposes the disentangled attention to model the importance of the single feature and interaction pairs, achieving the superior performance. However, these methods focus on enhancing the general feature interaction modeling ability for all features without sufficiently distinguishing valuable user features to improve user modeling.

In this paper, we propose a novel bi-level user modeling framework designed to enhance various DRS backbones. We present results from combining this framework with basic MLP, attention-based methods (DESTINE \cite{xu2021destine}), cross networks (DCN \cite{wang2017dcn}), gating methods (GDCN \cite{wang2023GDCN}), and the latest structure, FinalMLP \cite{mao2023finalmlp}.

\textbf{User Modeling.} User modeling methods can be categorized based on the modeling level: individual modeling and group modeling. In individual modeling, APG \cite{yan2022apg} and PPNet \cite{chang2023ppnet} typically generate personalized parameters based on user features to capture fine-grained user patterns and integrate these parameters for embeddings or prediction MLP. Conversely, group modeling aims to identify shared patterns within user groups, usually including the group division and pattern modeling. A straightforward approach involves categorizing users based on specific attributes and applying multi-task and multi-domain frameworks \cite{ma2018mmoe,tang2020ple,misra2016cross,sheng2021star} to model the group pattern. Recently, DGPM \cite{ma2023JD} provides a flexible group division method and feeds the group parameters to the prediction MLP for group modeling. It assigns group embeddings and prediction parameters to each group, which are updated by a Memory Network \cite{graves2014memorynet}. Users are first categorized into groups based on their similarities to group embeddings. Then, corresponding prediction parameters are feed to the prediction MLP for group-wise recommendations.

While existing user modeling approaches typically concentrate on a single level, either group or individual, \name introduces a novel approach by integrating bi-level modeling. Our approach has demonstrated superior performance in recommendation tasks.

%% file: 5Con.tex
\begin{figure}[t]
    \centering
    \includegraphics[width=0.95\linewidth]{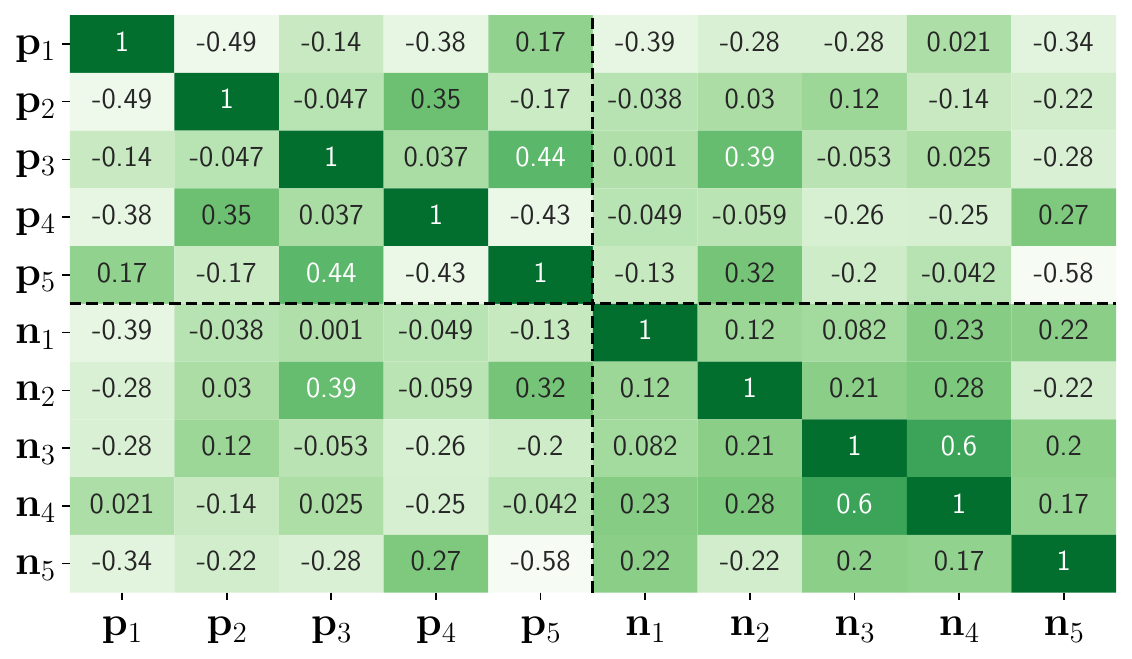}
    \vspace{-3mm}
    \caption{Similarity visualization for \name.}
    \label{fig:sim}
    \vspace{-3mm}
\end{figure}
\section{Conclusion}
In this study, we introduced \name, a cutting-edge approach to user modeling, which incorporates both group and individual levels and significantly improves deep recommender systems. \name innovatively utilizes a learnable classifier to categorize users based on user features and employs dual group embeddings to represent their group patterns. It then captures individual preferences from user personal features. These elements are seamlessly integrated into various recommendation models using flexible construction strategies for practical applications. Our comprehensive experiments have demonstrated \name's superiority.

\section*{Acknowledgments}
This research was partially supported by Research Impact Fund (No.R1015-23), APRC - CityU New Research Initiatives (No.9610565, Start-up Grant for New Faculty of CityU), CityU - HKIDS Early Career Research Grant (No.9360163), Hong Kong ITC Innovation and Technology Fund Midstream Research Programme for Universities Project (No.ITS/034/22MS), Hong Kong Environmental and Conservation Fund (No. 88/2022), and SIRG - CityU Strategic Interdisciplinary Research Grant (No.7020046), Huawei (Huawei Innovation Research Program), Tencent (CCF-Tencent Open Fund, Tencent Rhino-Bird Focused Research Program), Ant Group (CCF-Ant Research Fund, Ant Group Research Fund), Alibaba (CCF-Alimama Tech Kangaroo Fund (No. 2024002)), CCF-BaiChuan-Ebtech Foundation Model Fund, and Kuaishou.